\documentclass[sigconf]{acmart}

\AtBeginDocument{%
  \providecommand\BibTeX{{%
    \normalfont B\kern-0.5em{\scshape i\kern-0.25em b}\kern-0.8em\TeX}}}

\copyrightyear{2022}
\acmYear{2022}
\setcopyright{rightsretained}
\acmConference[WWW '22 Companion]{Companion Proceedings of the Web Conference 2022}{April 25--29, 2022}{Virtual Event, Lyon, France}
\acmBooktitle{Companion Proceedings of the Web Conference 2022 (WWW '22 Companion), April 25--29, 2022, Virtual Event, Lyon, France}\acmDOI{10.1145/3487553.3524653}
\acmISBN{978-1-4503-9130-6/22/04}

\begin{document}

\title{Assessing Network Representations for Identifying Interdisciplinarity}

\author{Eoghan Cunningham}
\orcid{0000-0002-0435-1962}
\affiliation{%
  \institution{School of Computer Science, University College Dublin}
  \country{Dublin, Ireland}
}
\email{eoghan.cunningham@ucdconnect.ie}
\author{Derek Greene}
\orcid{0000-0001-8065-5418}
\affiliation{%
  \institution{School of Computer Science, University College Dublin}
  \country{Dublin, Ireland}
}
\email{derek.greene@ucd.ie}

\renewcommand{\shortauthors}{Eoghan Cunningham \& Derek Greene}

\begin{abstract}
Many studies have sought to identify interdisciplinary research as a function of the diversity of disciplines identified in an article's references or citations. However, given the constant evolution of the scientific landscape, disciplinary boundaries are shifting and blurring, making it increasingly difficult to describe research within a strict taxonomy. In this work, we explore the potential for graph learning methods to learn embedded representations for research papers that encode their `interdisciplinarity' in a citation network. This facilitates the identification of interdisciplinary research without the use of disciplinary categories. We evaluate these representations and their ability to identify interdisciplinary research, according to their utility in interdisciplinary citation prediction. We find that those representations which preserve structural equivalence in the citation graph are best able to predict distant, interdisciplinary interactions in the network, according to multiple definitions of citation distance. 
\end{abstract}



\maketitle

\section{Introduction}
\emph{Interdisciplinary research} is most commonly defined as research activity which integrates
expertise, data or methodologies from two or more distinct disciplines. In light of its perceived importance, many studies have been conducted that quantify article and author interdisciplinarity in an effort to identify relevant research trends and to explore their impact. The most widely-accepted methods for measuring interdisciplinarity assess \emph{knowledge integration} using citation information, attempting to measure interdisciplinarity as a function of the balance, diversity, and dissimilarity of the topics identified in an article's cited papers \citep{rafols2010diversity,porter2007measuring, okamura2019interdisciplinarity}. Using these methods, a trend towards increased interdisciplinarity has been identified \citep{porter2009science}, which has been been found to correlate positively with research impact \citep{okamura2019interdisciplinarity,lariviere2015longdistance} and productivity \citep{leahey2017prominent}. 

Despite some agreement across measures of interdisciplinarity (with many studies based on work by \cite{porter2007measuring}), the literature lacks convergence towards a single metric \citep{wagner2011idr}. Most candidate metrics are reliant on explicit research topic or subject category information, for which sources are numerous, diverse, and sometimes unavailable. Moreover, the scientific community structure has been shown to be evolving rapidly \cite{rosvall2008maps}. This evolution may explain the lack of agreement between subject taxonomies and hierarchies, which vary considerably \cite{milojevic20science}.
The limitations of prescribed, static, subject classifications necessitate a modern approach that identifies research disciplines -- and indeed interdisciplinarity -- according to research network structure. Such an approach may offer a dynamic view of interdisciplinarity, and may help to map emerging developments that do not fit any existing schema \cite{wagner2011idr}. 

In this work, we begin to investigate appropriate graph representation learning methods which allow us to identify and quantify research interdisciplinarity in citation networks, so that we can monitor interdisciplinary interactions and track the development of research topics. We propose to learn unsupervised node (or paper)-level representations in research citation networks that encode research `interdisciplinarity'. In pursuit of such a representation, we explore the different network characteristics that embeddings must preserve in order to encode interdisciplinarity and we evaluate these embeddings according to their utility in an interdisciplinary citation prediction task. Specifically, we investigate how the performance of link prediction models (based on different node representations) is affected by the distance of citation, according to the hypothesis that representations that encode article interdisciplinarity will fare better when predicting ``far-reaching'' or ``long-distance'' interdisciplinary interactions/citations \cite{lariviere2015longdistance}, compared to those that do not. Given the acknowledged limitations of journal-level subject categories, we explore multiple definitions of citation distance: both according to prescribed categories with predefined topic distances, and to network defined distances according to graph structure.

\section{Background}

\subsection{Measuring Interdisciplinarity}

Numerous studies have developed metrics for quantifying research interdisciplinarity (IDR). Generally, these methods are designed to evaluate interdisciplinarity in pursuit of three goals: (i) to explore trends in IDR over time \citep{porter2009science}, (ii) to measure the benefits of IDR \citep{okamura2019interdisciplinarity,leahey2017prominent, lariviere2015longdistance}, (iii) to measure changes in interdisciplinarity in response to IDR initiatives \citep{porter2007measuring}.
While the proposed approaches are diverse, interdisciplinarity is most commonly measured on the basis of citation analysis. Using citations to model the flow of information, \emph{knowledge integration} is calculated as a function of the balance, diversity and/or distance between the disciplines identified in \emph{referenced} articles \citep{porter2009science,rafols2010diversity, okamura2019interdisciplinarity, abramo2018comparison}. Conversely, \emph{knowledge diffusion} is measured according to the disciplines of \emph{citing} articles \citep{van2015interdisciplinary,porter1985indicator}. These definition of knowledge integration, knowledge diffusion, or a combination of both have been used to enumerate interdisciplinarity for research papers. Invariably, such methods are reliant on explicit topic or discipline categories for research papers, such as those provided by Web of Science, Scopus, or Microsoft Academic. Such explicit categorisations for research papers, especially those assigned at a journal level, are problematic \cite{milojevic20science,abramo2018comparison}. Moreover, the inconsistencies evident across many of these subject taxonomies \cite{shen2019node2vec} may confirm that no singular, correct categorisation exists. However, recent graph learning methods may be capable of learning article representations that encode interdisciplinarity, without any knowledge of explicit discipline categorisation.

\subsection{Graph Learning - Node Embeddings}
Motivated by extensive work on embedded representations in natural language processing \citep{mikolov13efficient}, recently there has been considerable focus in the field of network analysis on developing analogous approaches for transforming nodes into low-dimensional vector representations \citep{cai18survey,goyal2018graph}. Methods for producing such representations, or node embeddings, are designed to preserve important relationships between nodes on the graph. In particular, we focus on approaches which preserve \emph{proximity} (where similar embeddings are learned for nodes which belong to the same community in the graph) and \emph{structural equivalence} (where similar embeddings are learned for nodes which have the same structural role in the graph). 

The \emph{DeepWalk} method \cite{perozzi2014deepwalk}, which extends the natural language processing-inspired SkipGram model \citep{mikolov13efficient}, generates fixed-length random walks from all nodes on the graph. The model learns to predict the nodes that will occur on walks given the starting node as input. As such, the model preserves proximities between nodes by learning similar embeddings for nodes which occur in similar contents, i.e. nodes that co-occur on random walks. \emph{node2vec}, proposed by \cite{grover2016node2vec}, extends DeepWalk, introducing two parameters which can be used to bias the random walks in order to explore different aspects of network structure. In our experiments, we implement both DeepWalk and node2vec as examples of proximity preserving node embeddings.
We also explore the structural equivalence preserving `role embeddings' learned via \emph{role2vec} \cite{ahmed2019role2vec}. The role2vec model further generalises the DeepWalk model attributing structural features (e.g. graph motif counts) to nodes before mapping those nodes to `roles' and conducting role-based random walks. 

\section{Data}
We conduct experiments using five citation networks, where each paper is represented as a node and citations are represented as undirected edges between them. Three of these datasets are well-known benchmark networks used to evaluate graph learning algorithms. We also contribute two new datasets which we have collected. 

\subsection{Benchmark Datasets}
\label{sec:benchmark_dataset_description}
As our three benchmark bibliographic datasets we consider \emph{Cora} \cite{mccallum00automating}, \emph{CiteSeer} \cite{qing03cite}, and \emph{PubMed} \cite{namata12pubmed}. These datasets range from 2,708 to 19,717 nodes (papers), with an average node degree (citations per paper) ranging from 2.7 -- 4.5. Each paper is assigned a label, representing a topic or field within a discipline. For example, the Cora dataset contains 7 topics within the field of Computer Science: \emph{`case based'}, \emph{`genetic algorithms'}, \emph{'neural networks'}, \emph{`probabilistic methods'}, \emph{`reinforcement learning'}, \emph{`rule learning'}, and \emph{`theory'}. While these datasets offer an accessible medium for exploring research networks, the insights gained from them may be limited by their sparsity (low node degree) and the scale of interdisciplinarity, i.e. they represent interactions between topics within a discipline, rather than macro interdisciplinary interactions. As such, we also provide two additional citation graphs that are both more complete and more interdisciplinary in nature. 

\subsection{Scopus Indexed Datasets}
We construct new citation networks using Microsoft Academic Graph citation data from two different sets of journal papers. These sets consist of two samples of articles from Scopus indexed journals, stratified according to their All Science Journal Categories (ASJC). We define two networks containing different topics, so that we can investigate interactions between different subsets of disciplines. We refer to these networks as \emph{Scopus1} and \emph{Scopus2}. The former contains samples of 1,500 articles published between 2017 and 2018 in Scopus indexed journals with the ASJCs \emph{`Computer Science', `Mathematics', `Medicine', `Chemistry'}, and \emph{`Social Sciences'}. For Scopus2, we use the seed set of categories: \emph{`Computer Science', `Mathematics', `Neuroscience', `Engineering', } and \emph{`Biochemistry, Genetics and Molecular Biology'}. For both datasets, we maximise the completeness of the graph by including all available referenced articles that are published in Scopus indexed journals. In this manner, we produce dense, multidisciplinary citation networks, such that each article can be categorised according to the ASJC of the journal in which it was published. Scopus1 contains 27,213 nodes with average degree 6.9, while Scopus2 contains 25,961 nodes with average degree 6.2. See statistics for all datasets in Table 1 of the Appendix. 

\section{Methods}

\subsection{Citation Prediction}

One established method for evaluating the quality of node embeddings is to assess their performance in a down-stream link prediction task \citep{kipf2016variational,grover2016node2vec}. We introduce the following citation prediction task to evaluate the quality of our paper representations and their ability to encode the citation graph structure. Further, we explore their ability to identify and predict interdisciplinary interactions in the network through their performance in interdisciplinary citation prediction.
We evaluate citation prediction performance with different node representations using 75:5:20 train:validation:test splits of the edges in each network. Three sets of node representations (DeepWalk, node2vec, and role2vec) are learned using the subgraph induced by the training edges.  In the case of each model, we learn node embeddings with 128 dimensions. To train a Multi-Layer Perception (MLP) model for the citation prediction task, we supplement each set of edges with a set of negative edges that were not present in the graph. We then represent each edge as the concatenation of the embedded representations of the associated nodes, and use train and validation edge sets to train an MLP model to classify positive and negative edges. We report performance on the test data as the area under the Receiver Operating Characteristic curve (AUC). Each experiment is repeated for 5 random splits of the data and the average AUC score is reported. Using this approach we evaluate the utility of different node representations in both overall citation prediction and in interdisciplinary citation prediction. 

\begin{figure}[h]
  \centering
  \includegraphics[width=0.7\linewidth]{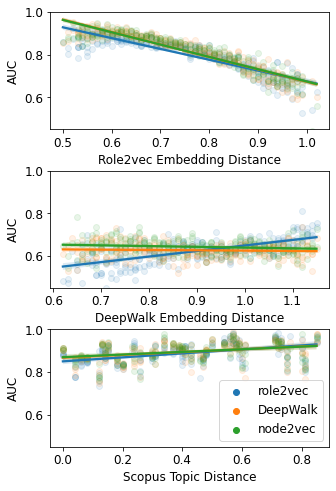}
  \caption{\emph{Scopus1} prediction performance (AUC) plotted against citation distance, for different node representations and different metrics of citation distance.}
  \label{fig:topic_distance_auc}
\end{figure}

We use the following methods to identify interdisciplinary citations in the test data according to both binary and continuous definitions of interdisciplinary. We report the binary interdisciplinary citation prediction score as the AUC performance on the test data for only the edges where the associated nodes belong to different classifications. These scores appear in parentheses in Table \ref{tab:benchmark_link_prediction}. In Figure \ref{fig:topic_distance_auc} we compare the relationship between citation prediction performance and citation distance according to our continuous definitions of interdisciplinarity, according to the intuition that representations that encode article interdisciplinarity will fare better when predicting ``long-distance'' interdisciplinary citations. To explore this relationship, we aggregate the edges in the test data (into bins of equal width) according to the chosen definition of citation distance, and plot the AUC for bins that contain at least 25 positive and negative edges. 

\subsection{Citation Distance Metrics}

To allow us to define a continuous form of interdisciplinarity, we implement and evaluate five definitions of citation distance: \emph{Scopus Topic Distance}, \emph{Network Distance}, \emph{DeepWalk Embedding Distance}, \emph{Node2vec Embedding Distance}, and \emph{Role2vec Embedding Distance}.
Scopus Topic Distance is defined according the neighbourhood similarity of the ASJC. Using a network of 6,000,000 papers published between 2010 and 2020, we create a discipline-level weighted adjacency matrix $\mathbf{W}$, that encodes all of the citations between categories in the ASJC. We define the topic distance between two categories $i$ and $j$ as the value $1 - S_{ij}$, where $S_{ij}$ is the cosine-similarity between rows $i$ and $j$ in $\mathbf{W}$. Thus, we can calculate the Scopus Topic Distance of a citation as the topic distance between the ASJC assigned to each of the relevant papers. 
We define the Network Distance of a citation as the length of the shortest path on the graph between to the two relevant papers \cite{wasserman94sna}. Finally, each of the embedding distances is defined as the cosine distance between the two vectors learned for the pair of the papers involved in the citation. 

\section{Results}

\subsection{Citation Prediction and Binary Interdisciplinarity}

\subsubsection{Benchmark Datasets}

Table \ref{tab:benchmark_link_prediction} shows citation results for 3 link prediction models, evaluated on 5 citation datasets. In each case, the AUC score is reported for both the entire test set and the interdisciplinary portion of the test set (IDR AUC). For each dataset, the structural representations learned by role2vec outperform those learned by DeepWalk-based methods. Although predicting interdisciplinary citations may be slightly more difficult than predicting intradisciplinary citations, performance in each task seems to correlate well. That is, the models which perform best at predicting citations within topics are also the most performant when predicting citations between topics. However, due to the limitations in scale and completeness of the benchmark datasets (see Section \ref{sec:benchmark_dataset_description}), we continue experiments using two new citation graphs collected from journals indexed by Scopus. 

\begin{table}
\caption{Citation prediction results: AUC (IDR AUC).}
\vskip -0.5em
\begin{tabular}{p{1.4cm}rrr}
\toprule
{Dataset} &   node2vec \cite{grover2016node2vec} & DeepWalk \cite{perozzi2014deepwalk} & role2vec \cite{ahmed2019role2vec} \\
\midrule
\emph{Cora}  &  0.785 (0.789) &  0.782 (0.769) & 0.836 (0.817)  \\
\emph{CiteSeer}  & 0.700 (0.670)  &    0.692 (0.644) &   0.766 (0.750)\\
\emph{PubMed}  &  0.800 (0.788) &   0.800 (0.778)  &  0.823 (0.818) \\
\emph{Scopus1} &  0.909 (0.903) &  0.906 (0.901) &    0.900 (0.896) \\
\emph{Scopus2} &  0.879 (0.875) &  0.879 (0.874) &  0.868 (0.869) \\
\bottomrule
\end{tabular}
\label{tab:benchmark_link_prediction}
\end{table}

\subsubsection{Scopus Indexed Datasets}

Table \ref{tab:benchmark_link_prediction} reports citation prediction performance in relation to the Scopus datasets. In the case of these larger, more diverse and denser networks, we find link prediction models are more successful, likely due to the greater completeness of the graph. Moreover, the DeepWalk-based methods are no longer less effective than role2vec embeddings, 
and the gap between intradisciplinary and interdisciplinary scores has reduced further.
To understand better the ability of different representations to encode the interdisciplinary role of research articles (by identifying those which can that can predict citations at greater distances), we now move to a continuous definition of citation distance.

\subsection{Citation Distance Metrics}

In our Scopus indexed networks, we explore 5 definitions of citation distance: 1 according to journal-level ASJC classifications and their citation neighbourhoods, 3 according to the distance between different unsupervised embedding representations, and 1 according to shortest path distance between papers. We compare these definitions of citation distance below, where we evaluate their relationship with network modularity and edge betweenness. 

\subsubsection{Edge Distance and Modularity}
We compare frequency distributions for distances of positive and negative edges to examine the extent to which different metrics of edge distance capture modularity in the Scopus networks. See the supplemental figures in the Appendix. Crucially, we find that the DeepWalk and node2vec defined citation distances offer the best separation between distributions of positive and negative edges. 

\subsubsection{Edge Distance and Edge Betweenness}
\label{sec:betweenness}

Table \ref{tab:edge_betweenness} shows Spearman rank correlations between citation distance and edge betweenness centrality, for each metric of citation distance in the Scopus indexed graph. The edge betweenness centrality for an edge (or citation) is a measure of the proportion of shortest paths on the graph which traverse that edge \cite{girvan20community}. So citations which bridge or connect communities will have high edge betweenness centrality. As such, we expect a definition of citation distance which encodes the interdisciplinarity of citation, to correlate positively with edge betweenness. The distance measures defined according to the distance between DeepWalk-based embeddings are shown to have the strongest correlation with edge betweenness centrality. Conversely, the distances between prescribed topics based on the ASJC from Scopus show no correlation with edge betweenness, indicating that these topic assignments, or at least the defined distances between them, do not reflect the community structure in the graph. 

We can further investigate the validity of the ASJC paper categorisations in the Scopus datasets, by comparing the edge betweenness scores of inter- and intra-topic edges. If the ASJC assignments represent the communities in the network, we would expect edges that are \emph{between} ASJC topics (i.e. citations linking papers of different categories) to have greater betweenness than those edges that are \emph{within} ASJC topics (i.e. linking papers with the same category). A Kolmogorov-Smirnov test comparing the edge betweenness distributions of inter- and intra-disciplinary edges finds interdisciplinary edges to have greater betweenness at 5\% significance. This indicates that, while the predefined distances between ASJC topics do not correlate with citation distance, the ASJC categorisations are somewhat reflective of the communities in the network.                                   

\begin{table}
\caption{Citation distances correlated with edge betweenness.}
\vskip -0.5em
\label{tab:edge_betweenness}
\begin{tabular}{p{2cm}rr}
\toprule
Distance&      \emph{Scopus1} & \emph{Scopus2} \\
\midrule
Scopus Topic      &  0.029 &  -0.020 \\
DeepWalk   &  0.331 &   0.300 \\
Node2vec &  0.329 &   0.304 \\
Role2vec &  0.191 &   0.193 \\
\bottomrule
\end{tabular}

\end{table}

\begin{table}
\caption{Link prediction AUC correlated with citation distance.}
\vskip -0.5em
\begin{tabular}{lrrr}
\toprule
Distance & role2vec & DeepWalk & node2vec \\
\midrule
Scopus Topic       &   0.444* &   0.327* &   0.323* \\
DeepWalk Embedding &   0.642* &   -0.061 &  -0.139* \\
Node2vec Embedding &   0.632* &  -0.126* &  -0.129* \\
Role2vec Embedding &  -0.875* &  -0.929* &  -0.914* \\
Network Distance            &    0.245 &   -0.173 &   -0.191 \\
\bottomrule
\end{tabular}
\label{tab:link_prediction_regression_1}
\end{table}

\subsubsection{Citation Distance and Link Prediction}
\label{sec:citation_distance_and_link_prediction}

Table \ref{tab:link_prediction_regression_1} reports the linear regression coefficients comparing citation prediction performance with citation distance in the \emph{Scopus1} network. The table include stars to highlight regression coefficients significant at 5\% significance. In particular, we note that the models employing role2vec structural representations of the nodes perform better as citation distance increases according to most definitions of citation distance. Table 5 in the Appendix plots similar results for \emph{Scopus2}, where role2vec performance again increases with those distances that agree with the underlying network structure. 

\section{Discussion and Conclusions}

In our investigations of the potential for graph learning methods to identify interdisciplinary research, we explored three different methods of node representation and evaluated their utility in predicting interdisciplinary citations. According to a binary definition of interdisciplinary citation, we found that predicting links between research topics is indeed more difficult than predicting within-topic interactions. This effect is more evident in sparser networks, such as the benchmark citation graphs \emph{Cora}, \emph{CiteSeer}, and \emph{PubMed}. 
In the more complete citation networks that we have collected (\emph{Scopus1} and \emph{Scopus2}), the distinction between citation prediction and interdisciplinary citation prediction performances are less apparent. 

We explored a continuous definition of interdisciplinary citation, as some  interactions are viewed as being more distant than others \cite{lariviere2015longdistance}. We evaluated five different definitions of citation distance. Although it remains difficult to determine the most accurate means of measuring the distance or interdisciplinarity of citation, we highlighted a number of issues with the predefined ASJC distances, which do not appear to agree with the underlying network structures. Further, we found that the role2vec-based model achieved citation prediction AUC that correlates positively with all metrics of citation distance that do appear to encode the underlying network structure. This provides evidence that the role2vec-based paper representations are capable of encoding the structural signatures associated with interdisciplinary interactions. We suggest that the embeddings which preserve structural equivalence on the graph are capable of encoding the `interdisciplinary role' of different articles.

We plan to extend this work to learning more advanced paper representations by including text features on nodes in the graph. 
There is also potential to develop means of better explaining these embeddings to quantify interdisciplinarity and interpret the different interdisciplinary roles of research papers.

\vskip 0.5em
\noindent\textbf{Acknowledgements.} This research was supported by Science Foundation Ireland (SFI) under Grant Number SFI/12/RC/2289\_P2. 

\bibliographystyle{ACM-Reference-Format}
\bibliography{main}


\begin{thebibliography}{25}


\ifx \showCODEN    \undefined \def \showCODEN     #1{\unskip}     \fi
\ifx \showDOI      \undefined \def \showDOI       #1{#1}\fi
\ifx \showISBNx    \undefined \def \showISBNx     #1{\unskip}     \fi
\ifx \showISBNxiii \undefined \def \showISBNxiii  #1{\unskip}     \fi
\ifx \showISSN     \undefined \def \showISSN      #1{\unskip}     \fi
\ifx \showLCCN     \undefined \def \showLCCN      #1{\unskip}     \fi
\ifx \shownote     \undefined \def \shownote      #1{#1}          \fi
\ifx \showarticletitle \undefined \def \showarticletitle #1{#1}   \fi
\ifx \showURL      \undefined \def \showURL       {\relax}        \fi
\providecommand\bibfield[2]{#2}
\providecommand\bibinfo[2]{#2}
\providecommand\natexlab[1]{#1}
\providecommand\showeprint[2][]{arXiv:#2}

\bibitem[\protect\citeauthoryear{Abramo, D’Angelo, and Zhang}{Abramo
  et~al\mbox{.}}{2018}]%
        {abramo2018comparison}
\bibfield{author}{\bibinfo{person}{Giovanni Abramo},
  \bibinfo{person}{Ciriaco~Andrea D’Angelo}, {and} \bibinfo{person}{Lin
  Zhang}.} \bibinfo{year}{2018}\natexlab{}.
\newblock \showarticletitle{A comparison of two approaches for measuring
  interdisciplinary research output: The disciplinary diversity of authors vs
  the disciplinary diversity of the reference list}.
\newblock \bibinfo{journal}{\emph{Journal of Informetrics}}
  \bibinfo{volume}{12}, \bibinfo{number}{4} (\bibinfo{year}{2018}),
  \bibinfo{pages}{1182--1193}.
\newblock


\bibitem[\protect\citeauthoryear{Ahmed, Rossi, Lee, Willke, Zhou, Kong, and
  Eldardiry}{Ahmed et~al\mbox{.}}{2019}]%
        {ahmed2019role2vec}
\bibfield{author}{\bibinfo{person}{Nesreen~K. Ahmed}, \bibinfo{person}{Ryan~A.
  Rossi}, \bibinfo{person}{John~Boaz Lee}, \bibinfo{person}{Theodore~L.
  Willke}, \bibinfo{person}{Rong Zhou}, \bibinfo{person}{Xiangnan Kong}, {and}
  \bibinfo{person}{Hoda Eldardiry}.} \bibinfo{year}{2019}\natexlab{}.
\newblock \showarticletitle{role2vec: Role-based network embeddings}. In
  \bibinfo{booktitle}{\emph{Proc. DLG KDD}}. \bibinfo{pages}{1--7}.
\newblock


\bibitem[\protect\citeauthoryear{Cai, Zheng, and Chang}{Cai
  et~al\mbox{.}}{2018}]%
        {cai18survey}
\bibfield{author}{\bibinfo{person}{Hongyun Cai}, \bibinfo{person}{Vincent~W
  Zheng}, {and} \bibinfo{person}{Kevin Chen-Chuan Chang}.}
  \bibinfo{year}{2018}\natexlab{}.
\newblock \showarticletitle{A comprehensive survey of graph embedding:
  Problems, techniques, and applications}.
\newblock \bibinfo{journal}{\emph{IEEE Transactions on Knowledge and Data
  Engineering}} \bibinfo{volume}{30}, \bibinfo{number}{9}
  (\bibinfo{year}{2018}), \bibinfo{pages}{1616--1637}.
\newblock


\bibitem[\protect\citeauthoryear{Girvan and Newman}{Girvan and Newman}{2002}]%
        {girvan20community}
\bibfield{author}{\bibinfo{person}{Michelle Girvan} {and}
  \bibinfo{person}{Mark~E.J. Newman}.} \bibinfo{year}{2002}\natexlab{}.
\newblock \showarticletitle{Community structure in social and biological
  networks}.
\newblock \bibinfo{journal}{\emph{Proceedings of the national academy of
  sciences}} \bibinfo{volume}{99}, \bibinfo{number}{12} (\bibinfo{year}{2002}),
  \bibinfo{pages}{7821--7826}.
\newblock


\bibitem[\protect\citeauthoryear{Goyal and Ferrara}{Goyal and Ferrara}{2018}]%
        {goyal2018graph}
\bibfield{author}{\bibinfo{person}{Palash Goyal} {and} \bibinfo{person}{Emilio
  Ferrara}.} \bibinfo{year}{2018}\natexlab{}.
\newblock \showarticletitle{Graph embedding techniques, applications, and
  performance: A survey}.
\newblock \bibinfo{journal}{\emph{Knowledge-Based Systems}}
  \bibinfo{volume}{151} (\bibinfo{year}{2018}), \bibinfo{pages}{78--94}.
\newblock


\bibitem[\protect\citeauthoryear{Grover and Leskovec}{Grover and
  Leskovec}{2016}]%
        {grover2016node2vec}
\bibfield{author}{\bibinfo{person}{Aditya Grover} {and} \bibinfo{person}{Jure
  Leskovec}.} \bibinfo{year}{2016}\natexlab{}.
\newblock \showarticletitle{node2vec: Scalable feature learning for networks}.
  In \bibinfo{booktitle}{\emph{Proc. 22nd ACM SIGKDD International Conference
  on Knowledge Discovery and Data Mining}}. \bibinfo{pages}{855--864}.
\newblock


\bibitem[\protect\citeauthoryear{Kipf and Welling}{Kipf and Welling}{2016}]%
        {kipf2016variational}
\bibfield{author}{\bibinfo{person}{Thomas~N Kipf} {and} \bibinfo{person}{Max
  Welling}.} \bibinfo{year}{2016}\natexlab{}.
\newblock \showarticletitle{Variational graph auto-encoders}.
\newblock \bibinfo{journal}{\emph{arXiv preprint arXiv:1611.07308}}
  (\bibinfo{year}{2016}).
\newblock


\bibitem[\protect\citeauthoryear{Larivi{\`e}re, Haustein, and
  B{\"o}rner}{Larivi{\`e}re et~al\mbox{.}}{2015}]%
        {lariviere2015longdistance}
\bibfield{author}{\bibinfo{person}{Vincent Larivi{\`e}re},
  \bibinfo{person}{Stefanie Haustein}, {and} \bibinfo{person}{Katy
  B{\"o}rner}.} \bibinfo{year}{2015}\natexlab{}.
\newblock \showarticletitle{Long-distance interdisciplinarity leads to higher
  scientific impact}.
\newblock \bibinfo{journal}{\emph{PloS one}} \bibinfo{volume}{10},
  \bibinfo{number}{3} (\bibinfo{year}{2015}),
  \bibinfo{pages}{e0122565--e0122565}.
\newblock


\bibitem[\protect\citeauthoryear{Leahey, Beckman, and Stanko}{Leahey
  et~al\mbox{.}}{2017}]%
        {leahey2017prominent}
\bibfield{author}{\bibinfo{person}{Erin Leahey}, \bibinfo{person}{Christine~M.
  Beckman}, {and} \bibinfo{person}{Taryn~L Stanko}.}
  \bibinfo{year}{2017}\natexlab{}.
\newblock \showarticletitle{Prominent but less productive: The impact of
  interdisciplinarity on scientists’ research}.
\newblock \bibinfo{journal}{\emph{Administrative Science Quarterly}}
  \bibinfo{volume}{62}, \bibinfo{number}{1} (\bibinfo{year}{2017}),
  \bibinfo{pages}{105--139}.
\newblock


\bibitem[\protect\citeauthoryear{Lu and Getoor}{Lu and Getoor}{2003}]%
        {qing03cite}
\bibfield{author}{\bibinfo{person}{Qing Lu} {and} \bibinfo{person}{Lise
  Getoor}.} \bibinfo{year}{2003}\natexlab{}.
\newblock \showarticletitle{Link-Based Classification}. In
  \bibinfo{booktitle}{\emph{Proc. 20th International Conference on
  International Conference on Machine Learning}} (Washington, DC, USA)
  \emph{(\bibinfo{series}{ICML'03})}. \bibinfo{publisher}{AAAI Press},
  \bibinfo{pages}{496–503}.
\newblock
\showISBNx{1577351894}


\bibitem[\protect\citeauthoryear{McCallum, Nigam, Rennie, and Seymore}{McCallum
  et~al\mbox{.}}{2000}]%
        {mccallum00automating}
\bibfield{author}{\bibinfo{person}{Andrew~Kachites McCallum},
  \bibinfo{person}{Kamal Nigam}, \bibinfo{person}{Jason Rennie}, {and}
  \bibinfo{person}{Kristie Seymore}.} \bibinfo{year}{2000}\natexlab{}.
\newblock \showarticletitle{Automating the construction of internet portals
  with machine learning}.
\newblock \bibinfo{journal}{\emph{Information Retrieval}} \bibinfo{volume}{3},
  \bibinfo{number}{2} (\bibinfo{year}{2000}), \bibinfo{pages}{127--163}.
\newblock


\bibitem[\protect\citeauthoryear{Mikolov, Chen, Corrado, and Dean}{Mikolov
  et~al\mbox{.}}{2013}]%
        {mikolov13efficient}
\bibfield{author}{\bibinfo{person}{Tomas Mikolov}, \bibinfo{person}{Kai Chen},
  \bibinfo{person}{Greg Corrado}, {and} \bibinfo{person}{Jeffrey Dean}.}
  \bibinfo{year}{2013}\natexlab{}.
\newblock \showarticletitle{Efficient estimation of word representations in
  vector space}.
\newblock \bibinfo{journal}{\emph{arXiv preprint arXiv:1301.3781}}
  (\bibinfo{year}{2013}).
\newblock


\bibitem[\protect\citeauthoryear{Milojevi{\'c}}{Milojevi{\'c}}{2020}]%
        {milojevic20science}
\bibfield{author}{\bibinfo{person}{Sta{\v s}a Milojevi{\'c}}.}
  \bibinfo{year}{2020}\natexlab{}.
\newblock \showarticletitle{{Practical method to reclassify Web of Science
  articles into unique subject categories and broad disciplines}}.
\newblock \bibinfo{journal}{\emph{Quantitative Science Studies}}
  \bibinfo{volume}{1}, \bibinfo{number}{1} (\bibinfo{date}{02}
  \bibinfo{year}{2020}), \bibinfo{pages}{183--206}.
\newblock


\bibitem[\protect\citeauthoryear{Namata, London, Getoor, Huang, and EDU}{Namata
  et~al\mbox{.}}{2012}]%
        {namata12pubmed}
\bibfield{author}{\bibinfo{person}{Galileo Namata}, \bibinfo{person}{Ben
  London}, \bibinfo{person}{Lise Getoor}, \bibinfo{person}{Bert Huang}, {and}
  \bibinfo{person}{UMD EDU}.} \bibinfo{year}{2012}\natexlab{}.
\newblock \showarticletitle{Query-driven active surveying for collective
  classification}. In \bibinfo{booktitle}{\emph{Proc. 10th International
  Workshop on Mining and Learning with Graphs}}, Vol.~\bibinfo{volume}{8}.
  \bibinfo{pages}{1}.
\newblock


\bibitem[\protect\citeauthoryear{Okamura}{Okamura}{2019}]%
        {okamura2019interdisciplinarity}
\bibfield{author}{\bibinfo{person}{Keisuke Okamura}.}
  \bibinfo{year}{2019}\natexlab{}.
\newblock \showarticletitle{Interdisciplinarity revisited: evidence for
  research impact and dynamism}.
\newblock \bibinfo{journal}{\emph{Palgrave Communications}}
  \bibinfo{volume}{5}, \bibinfo{number}{1} (\bibinfo{year}{2019}),
  \bibinfo{pages}{1--9}.
\newblock


\bibitem[\protect\citeauthoryear{Perozzi, Al-Rfou, and Skiena}{Perozzi
  et~al\mbox{.}}{2014}]%
        {perozzi2014deepwalk}
\bibfield{author}{\bibinfo{person}{Bryan Perozzi}, \bibinfo{person}{Rami
  Al-Rfou}, {and} \bibinfo{person}{Steven Skiena}.}
  \bibinfo{year}{2014}\natexlab{}.
\newblock \showarticletitle{Deepwalk: Online learning of social
  representations}. In \bibinfo{booktitle}{\emph{Proc. 20th ACM SIGKDD
  international conference on Knowledge discovery and data mining}}.
  \bibinfo{pages}{701--710}.
\newblock


\bibitem[\protect\citeauthoryear{Porter and Chubin}{Porter and Chubin}{1985}]%
        {porter1985indicator}
\bibfield{author}{\bibinfo{person}{ALDE Porter} {and} \bibinfo{person}{D
  Chubin}.} \bibinfo{year}{1985}\natexlab{}.
\newblock \showarticletitle{An indicator of cross-disciplinary research}.
\newblock \bibinfo{journal}{\emph{Scientometrics}} \bibinfo{volume}{8},
  \bibinfo{number}{3-4} (\bibinfo{year}{1985}), \bibinfo{pages}{161--176}.
\newblock


\bibitem[\protect\citeauthoryear{Porter, Cohen, David~Roessner, and
  Perreault}{Porter et~al\mbox{.}}{2007}]%
        {porter2007measuring}
\bibfield{author}{\bibinfo{person}{Alan Porter}, \bibinfo{person}{Alex Cohen},
  \bibinfo{person}{J. David~Roessner}, {and} \bibinfo{person}{Marty
  Perreault}.} \bibinfo{year}{2007}\natexlab{}.
\newblock \showarticletitle{Measuring researcher interdisciplinarity}.
\newblock \bibinfo{journal}{\emph{Scientometrics}} \bibinfo{volume}{72},
  \bibinfo{number}{1} (\bibinfo{year}{2007}), \bibinfo{pages}{117--147}.
\newblock


\bibitem[\protect\citeauthoryear{Porter and Rafols}{Porter and Rafols}{2009}]%
        {porter2009science}
\bibfield{author}{\bibinfo{person}{Alan Porter} {and} \bibinfo{person}{Ismael
  Rafols}.} \bibinfo{year}{2009}\natexlab{}.
\newblock \showarticletitle{Is science becoming more interdisciplinary?
  Measuring and mapping six research fields over time}.
\newblock \bibinfo{journal}{\emph{Scientometrics}} \bibinfo{volume}{81},
  \bibinfo{number}{3} (\bibinfo{year}{2009}), \bibinfo{pages}{719--745}.
\newblock


\bibitem[\protect\citeauthoryear{Rafols and Meyer}{Rafols and Meyer}{2010}]%
        {rafols2010diversity}
\bibfield{author}{\bibinfo{person}{Ismael Rafols} {and} \bibinfo{person}{Martin
  Meyer}.} \bibinfo{year}{2010}\natexlab{}.
\newblock \showarticletitle{Diversity and network coherence as indicators of
  interdisciplinarity: case studies in bionanoscience}.
\newblock \bibinfo{journal}{\emph{Scientometrics}} \bibinfo{volume}{82},
  \bibinfo{number}{2} (\bibinfo{year}{2010}), \bibinfo{pages}{263--287}.
\newblock


\bibitem[\protect\citeauthoryear{Rosvall and Bergstrom}{Rosvall and
  Bergstrom}{2008}]%
        {rosvall2008maps}
\bibfield{author}{\bibinfo{person}{Martin Rosvall} {and}
  \bibinfo{person}{Carl~T Bergstrom}.} \bibinfo{year}{2008}\natexlab{}.
\newblock \showarticletitle{Maps of random walks on complex networks reveal
  community structure}.
\newblock \bibinfo{journal}{\emph{Proceedings of the National Academy of
  Sciences}} \bibinfo{volume}{105}, \bibinfo{number}{4} (\bibinfo{year}{2008}),
  \bibinfo{pages}{1118--1123}.
\newblock


\bibitem[\protect\citeauthoryear{Shen, Chen, Yang, and Wu}{Shen
  et~al\mbox{.}}{2019}]%
        {shen2019node2vec}
\bibfield{author}{\bibinfo{person}{Zhesi Shen}, \bibinfo{person}{Fuyou Chen},
  \bibinfo{person}{Liying Yang}, {and} \bibinfo{person}{Jinshan Wu}.}
  \bibinfo{year}{2019}\natexlab{}.
\newblock \showarticletitle{Node2vec representation for clustering journals and
  as a possible measure of diversity}.
\newblock \bibinfo{journal}{\emph{Journal of Data and Information Science}}
  \bibinfo{volume}{4}, \bibinfo{number}{2} (\bibinfo{year}{2019}),
  \bibinfo{pages}{79}.
\newblock


\bibitem[\protect\citeauthoryear{Van~Noorden et~al\mbox{.}}{Van~Noorden
  et~al\mbox{.}}{2015}]%
        {van2015interdisciplinary}
\bibfield{author}{\bibinfo{person}{Richard Van~Noorden} {et~al\mbox{.}}}
  \bibinfo{year}{2015}\natexlab{}.
\newblock \showarticletitle{Interdisciplinary research by the numbers}.
\newblock \bibinfo{journal}{\emph{Nature}} \bibinfo{volume}{525},
  \bibinfo{number}{7569} (\bibinfo{year}{2015}), \bibinfo{pages}{306--307}.
\newblock


\bibitem[\protect\citeauthoryear{Wagner, Roessner, Bobb, Klein, Boyack, Keyton,
  Rafols, and B{\"o}rner}{Wagner et~al\mbox{.}}{2011}]%
        {wagner2011idr}
\bibfield{author}{\bibinfo{person}{Caroline~S. Wagner},
  \bibinfo{person}{J.~David Roessner}, \bibinfo{person}{Kamau Bobb},
  \bibinfo{person}{Julie~Thompson Klein}, \bibinfo{person}{Kevin~W. Boyack},
  \bibinfo{person}{Joann Keyton}, \bibinfo{person}{Ismael Rafols}, {and}
  \bibinfo{person}{Katy B{\"o}rner}.} \bibinfo{year}{2011}\natexlab{}.
\newblock \showarticletitle{Approaches to understanding and measuring
  interdisciplinary scientific research (IDR): A review of the literature}.
\newblock \bibinfo{journal}{\emph{Journal of Informetrics}}
  \bibinfo{volume}{5}, \bibinfo{number}{1} (\bibinfo{year}{2011}),
  \bibinfo{pages}{14--26}.
\newblock


\bibitem[\protect\citeauthoryear{Wasserman and Faust}{Wasserman and
  Faust}{1994}]%
        {wasserman94sna}
\bibfield{author}{\bibinfo{person}{Stanley Wasserman} {and}
  \bibinfo{person}{Katherine Faust}.} \bibinfo{year}{1994}\natexlab{}.
\newblock \bibinfo{booktitle}{\emph{{Social Network Analysis: Methods and
  Applications}}}.
\newblock \bibinfo{publisher}{Cambridge University Press}.
\newblock


\end{thebibliography}
\clearpage
\appendix

\section{Supplementary Materials}

\subsection{Summary of Datasets}
Table \ref{tab:data_description} below includes descriptive statistics for the five citation datasets considered in our experiments, including the number of unique topics (categories) per dataset.

\begin{table}[!h]
\caption{Summary of datasets.}
\begin{tabular}{lrrr}
\toprule
{} &   Papers &  Citations per Paper &   Topics \\
Dataset\;\;\;\;\;\;\;  &           &                     &           \\
\midrule
\emph{Cora}     &      2,708 &                3.90 &         7 \\
\emph{CiteSeer} &      3,327 &                2.74 &         6 \\
\emph{PuMmed}   &     19,717 &                4.50 &         3 \\
\emph{Scopus1}   &     27,162 &                6.91 &        25 \\
\emph{Scopus2}  &     25,961 &                6.15 &        26 \\
\bottomrule
\end{tabular}
\label{tab:data_description}
\end{table}

\subsection{Scopus Edge Distance and Modularity}
\label{sec:modularity}

\begin{figure}[H]
  \centering
  \includegraphics[width=0.75\linewidth]{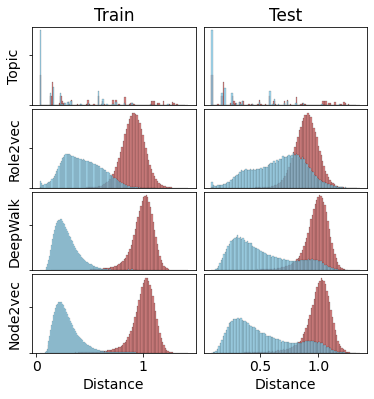}
  \caption{Edge distance distribution for positive (blue) and negative (red) edges in the \emph{Scopus1} dataset.}
  \Description{Edge distance distribution for positive (blue) and negative (red) edges in the Scopus Indexed dataset.}
    \label{fig:edge_distance_distributions_1}
\end{figure}

\begin{figure}[!t]
  \centering
  \includegraphics[width=0.75\linewidth]{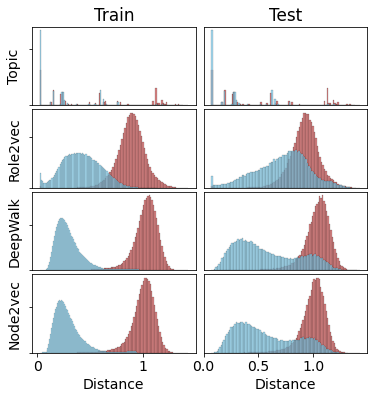}
  \caption{Edge distance distribution for positive (blue) and negative (red) edges in the \emph{Scopus2} dataset.}
  \Description{Edge distance distribution for positive (blue) and negative (red) edges in the Scopus Indexed dataset.}
    \label{fig:edge_distance_distributions_2}
\end{figure}

Figures \ref{fig:edge_distance_distributions_1} and \ref{fig:edge_distance_distributions_2} illustrate the edge distance distributions for both the positive and negative edges in the training and test portions of the Scopus indexed datasets. For each definition of edge distance, the frequency distribution for the distance of positive edges is plotted in blue, and the distance of negative edges is plotted in red. 
The DeepWalk-based methods (DeepWalk and node2vec) appear to learn node representations that encode the emergent communities in the graph, as adjacent nodes are assigned similar/near embeddings, while non-adjacent nodes (represented by the negative edge distribution in red) are found to have more distant embeddings.

\begin{table}[!t]
\caption{Link prediction AUC scores correlated with citation distances.}
\begin{tabular}{lrrr}
\toprule
Distance & role2vec & DeepWalk & node2vec \\
\midrule
Scopus Topic       &   -0.036 &   -0.006 &    0.059 \\
DeepWalk Embedding &   0.144* &  -0.456* &  -0.373* \\
Node2vec Embedding &   0.284* &  -0.357* &  -0.218* \\
Role2vec Embedding &  -0.843* &  -0.778* &  -0.807* \\
Network Distance            &   0.296* &  -0.388* &  -0.401* \\
\bottomrule
\end{tabular}
\label{tab:link_prediction_regression_2}
\end{table}

\begin{figure}[H]
  \centering
  \includegraphics[width=0.75\linewidth]{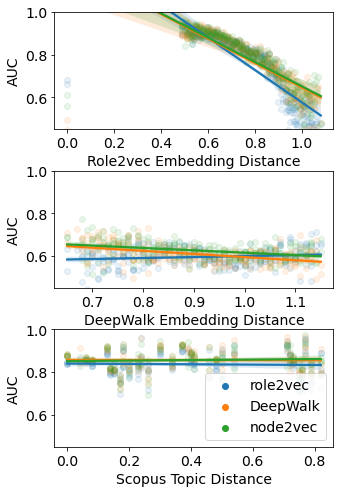}
  \caption{\emph{Scopus2} citation prediction performance (AUC) plotted against citation distance, for different node representations and different metrics of citation distance.}
  \Description{Edge distance distribution for positive (blue) and negative (red) edges in the Scopus Indexed dataset.}
    \label{fig:topic_distance_auc_2}
\end{figure}

\end{document}